\documentclass[twocolumn,twoside,slac_two]{revtex4}
\usepackage{graphicx}
\usepackage{fancyhdr}
\begin{document}

\title{Improved Factorization Method in Studying
$B$-meson Decays}
\author{Marina--Aura Dariescu and Ciprian Dariescu}
\affiliation{Department of Solid State and Theoretical Physics,
Al. I. Cuza University, Ia\c{s}i, Romania}

\begin{abstract}
$B$ decays are a subject of active research since they provide
useful information on the dynamics of strong and electroweak
interactions for testing the Standard Model (SM) and models beyond
and are ideally suited for a critical analysis of CP violation
phenomena. Within the standard model, there exist certain
relations between CP violating rate differences in $B$ decays in
the $SU(3)$ limit, as for example $\Delta (\bar{B}^0 \to \pi^+
\pi^-) = - \Delta (\bar{B}^0 \to \pi^+ K^-)$. The goal of this
letter is to study the direct CP violation asymmetry in a class of
processes where there has been recent theoretical progress, as for
example the $B$ decays into two light pseudoscalars mesons and
into a light pseudoscalar and a light vector meson. We identify
relations between rate asymmetries which are valid in the SU(3)
limit in the standard model and we compute SU(3) breaking
corrections to them, going beyond the naive factorization by using
the QCD improved factorization model of Beneke {\it et al.}.
Finally, in some processes as for example $BR(B^- \to
\eta^{\prime} K^-)$, we claim that one has to add SUSY
contributions to the Wilson coefficients. In these cases, we end
with a $BR$ depending on three parameters, whose values are
constrained by the experimental data.
\end{abstract}

\maketitle

\thispagestyle{fancy}

\section{INTRODUCTION}

As it is known, in the Standard Model, the CP violation arises
solely from the phase in the $3 \times 3$ unitary CKM matrix and
any CP violating observable is proportional to
$Im(V_{ij}V_{il}^*V_{kj}^*V_{kl})$, with $i\neq k$ and $j\neq l$.
The SU(3) invariant amplitude for $B \to PP$ and $B \to PV$
decays, in terms of the tree and penguin contributions, are, for
example,
\begin{eqnarray}
A(\bar{B}^0 \to \pi^+ \pi^- ) & = & V_{ub} V_{ud}^* \, T + V_{cb}
V_{cd}^* \, P \, , \nonumber \\* A(\bar{B}^0 \to \pi^+ K^- ) & = &
V_{ub} V_{us}^* \, T + V_{cb} V_{cs}^* \, P \, , \nonumber
\end{eqnarray}
with $T$ and $P$ the same in the two processes. Even there are no
simple relations among the branching ratios of these decays since
the CKM factors in the $T$ and $P$ amplitudes are different, one
has the following relations among the CP violating rate
differences, [6],
\begin{eqnarray}
\Delta^{\bar B^0}_{\pi^+\pi^-} = - \Delta^{\bar B_s^0}_{K^+ K^-} =
\Delta^{\bar B^0_s}_{K^+ \pi^-} = - \Delta^{\bar B^0}_{\pi^+ K^-}
\, ,
\end{eqnarray}
and similarly for $B \to PV$, where
\begin{equation}
\Delta^B_{PP} = \Gamma(B\to PP) - \bar \Gamma(\bar B \to \bar
P\bar P) \, ,
\end{equation}
while the CP asymmetry is defined as
\begin{equation}
A_{CP} = \frac{\Gamma(B\to PP) - \bar \Gamma(\bar B \to \bar P\bar
P)}{ \Gamma(B\to PP) + \bar \Gamma(\bar B \to \bar P\bar P)} \,  .
\end{equation}

The most important question now is to establish to what precision
these relations hold within the standard model, or equivalently to
estimate the corrections they might receive from different
sources, as for example the SU(3)breaking effects.

In this respect, let us discuss the corrections in {\it QCD
improved factorization method} developed by Beneke et al. [3, 4]
and compare the results with the ones obtained within the so
called {\it Naive Factorization}, [1], which will be briefly
presented in the next section.
\\

\section{NAIVE FACTORIZATION}

The effective weak Hamiltonian for $B \to PP$ decays,[1],
\begin{widetext}
\begin{equation}
H_{eff} = \frac{G_F}{\sqrt{2}} \sum_{p=u,c} \lambda_p \left[ C_1
O_1^p + C_2 O_2^p + \sum_{i=3}^{10} C_i O_i + C_{7 \gamma} O_{7
\gamma} + C_{8g} O_{8g} \right] + h.c.
\end{equation}
\end{widetext}
where $\lambda_p = V_{pb} V_{pq}^*$, with $p=u,c$ and $q=d$ (for
$\Delta S=0$ processes) and $q=s$ (for $\Delta S=1$ processes), in
the case when $a_1^c=a_2^c=0$, becomes
\begin{widetext}
\begin{equation}
H_{eff} = \frac{G_F}{\sqrt{2}} \left[ \lambda_u \left( C_1 O_1^u +
C_2 O_2^u \right) + \lambda_p \left( \sum_{i=3}^{10} C_i O_i^p +
C_{7 \gamma} O_{7 \gamma} + C_{8g} O_{8g} \right) \right] + h.c.
\end{equation}
\end{widetext}
It is expressed in terms of the Wilson coefficients $C_i$
(evaluated at the renormalization scale $\mu =m_B$), the usual
tree level left-handed current-current operators, the QCD and
electroweak penguin operators, the electromagnetic and
chromomagnetic dipole operators.

Within {\it naive factorization}, one is able to put the matrix
element of the Hamiltonian in terms of the form factors and decay
constant of the meson which is factorized as
\begin{widetext}
\begin{equation}
\langle P_1 P_2 | H_{eff} |B \rangle = i \frac{G_F}{\sqrt{2}}
\lambda_p \left( \frac{1}{N_c} C_i + C_j \right) f_{P_2} (
m_B^2-m_1^2) F_0^{B \to P_1}(m_2^2) + ( 1 \leftrightarrow 2 ) \, ,
\end{equation}
\end{widetext}
where $N_c=3$ is the number of colors and we
introduce the usual combinations of the Wilson coefficients
\begin{eqnarray}
& & a_i \equiv C_i + \frac{1}{3} C_{i+1} \; ({\rm for} \; i=odd) ,
\nonumber \\* & & a_i \equiv C_i + \frac{1}{3} C_{i-1} \; ({\rm
for} \; i=even) \,  .
\end{eqnarray}

In the concrete case of the $\bar{B}_0 \to \pi^+ \pi^-$ and
$\bar{B}_0 \to \pi^+ K^-$ processes, the matrix elements
respectively are, [1],
\begin{eqnarray}
& & A(\bar{B}_0 \to \pi^+ \pi^-) = - \, i \frac{G_f}{\sqrt{2}}
f_{\pi} F^{B \to \pi}_0 (m_{\pi}^2) (m_B^2-m_{\pi}^2) \nonumber
\\* & & \times \lbrace v_u^d \, a_1 - v_t^d \left[ a_4+a_{10} +
r_{\pi} (a_6 + a_8) \right] \rbrace , \nonumber \\* & &
A(\bar{B}_0 \to \pi^+ K^-) = - \, i \frac{G_f}{\sqrt{2}} f_K F^{B
\to \pi}_0 (m_K^2) (m_B^2-m_{\pi}^2) \nonumber \\* & & \times
\lbrace v_u^s \, a_1 - v_t^s \left[ a_4+a_{10} + r_K (a_6 + a_8)
\right] \rbrace ,
\end{eqnarray}
with
\begin{eqnarray}
& & v_u^d = V_{ub} V_{ud}^* \, , \; v_c^d = V_{cb} V_{cd}^* \, ,
\nonumber \\ && v_u^s = V_{ub} V_{us}^* \, , \; v_c^s = V_{cb}
V_{cs}^*
\end{eqnarray}
and
\begin{equation}
r_{\pi (K)} = \frac{2 m_{\pi (K)}^2}{(m_b-m_u) (m_u+m_{d(s)})}
\approx 1.2 \, .
\end{equation}
Using the unitarity relations and introducing the notations:
\begin{eqnarray}
& & \frac{v_u^d}{v_c^d} \equiv - \, R_d \, e^{-i \gamma} \, , \;
\; \frac{v_u^s}{v_c^s} \equiv R_s \, e^{-i \gamma} \, , \nonumber
\\* & & \alpha \equiv a_4 + a_{10} + r (a_6+a_8) \, , \; \; \beta
\equiv a_1 + \alpha
\end{eqnarray}
the amplitudes (8) get the expressions
\begin{widetext}
\begin{eqnarray}
A(\bar{B}_0 \to \pi^+ \pi^-) & = & - \, i \frac{G_f}{\sqrt{2}}
f_{\pi} F^{B \to \pi}_0 (m_B^2-m_{\pi}^2) v_c^d \left[ - R_d \,
e^{-i \gamma} \beta + \alpha \right] ,\nonumber \\* A(\bar{B}_0
\to \pi^+ K^-) & = & - \,  i \frac{G_f}{\sqrt{2}} f_{K} F^{B \to
\pi}_0 (m_B^2-m_{\pi}^2) v_c^s \left[ R_s \, e^{-i \gamma} \beta +
\alpha \right] ,
\end{eqnarray}
which allow us to write down the CP violating amplitude difference
\begin{eqnarray}
\left( |A|^2 - | \bar{A} |^2 \right)_{B \to \pi \pi} & = &
\frac{G_F^2}{2} f_{\pi}^2 (F_0^{B \to \pi})^2 (m_B^2-m_{\pi}^2)^2
4 (v_c^d)^2 R_d \sin \gamma \; \delta_{\pi \pi} \, , \nonumber \\*
\left( |A|^2 - | \bar{A} |^2 \right)_{B \to \pi K} & = & - \,
\frac{G_F^2}{2} f_{K}^2 (F_0^{B \to \pi})^2 (m_B^2-m_{\pi}^2)^2 4
(v_c^s)^2 R_s \sin \gamma \; \delta_{\pi K} \, , \nonumber \\*
\end{eqnarray}
\end{widetext}
where the quantity $\delta = {\rm Re} (\beta) {\rm Im} (\alpha) -
{\rm Im} (\beta) {\rm Re} (\alpha)$ is the same.

Within SU(3) flavor symmetry, when phase space differences are
neglected, naive factorization yields the relation, [8],
\begin{equation}
\Delta^{\bar B^0}_{\pi^+ \pi^-} = - {f^2_\pi\over f^2_K}
\Delta^{\bar B^0}_{\pi^+ K^-} \, ,
\end{equation}
which can be used to test the SM CP violation, or to predict one
rate difference if the other one is known. In these assumptions,
the relation between the CP asymmetries,
\begin{equation}
A_{CP}(\pi^+ \pi^-) = - {f_\pi^2 \over f^2_K} {Br(\pi^+ K^-)\over
Br(\pi^+ \pi^-)} \, A_{CP}(\pi^+ K^-),
\end{equation}
for the reported CP branching ratios [Babar]
\[
Br(\pi^+ \pi^-) = (5.8 \pm 0.4 \pm 0.3)\times 10^{-6} \] \[
Br(\pi^+ K^-) = (19.4 \pm 0.6) \times 10^{-6} \; ,
\]
leads to the following result,
\[
A_{CP}(\pi^+ \pi^-) \approx - 2.2 \, A_{CP} (\pi^+ K^-) \, ,
\]
which does not agree with the experimental data that are just
emerging from Babar, \[ A_{CP}(\pi^+ K^-) = -0.107 \pm 0.018 \]
and Belle, \[ A_{CP}(\pi^+ \pi^-) = 0.55 \pm 0.08 \pm 0.05 . \]

\section{IMPROVED FACTORIZATION METHOD}

Let us turn to the improved factorization method, (IFM), developed
by Beneke et al. [3], which gives a systematic and
model-independent calculation of two-body hadronic decays, in the
heavy-quark limit. The factorization formula, presented in the
previous section, is applicable, since the nonfactorizable
corrections are included in the $a_i$ parameters, which have
imaginary parts coming from vertex corrections and penguin
contributions. In this approach, the Wilson coefficients are
calculated at the scale $\mu = m_b$ using next-to-leading order
modified scheme, the electroweak penguin contributions are
considered as next-to-leading order and there is also a
contribution coming from the hard scattering with the spectator.

The IFM formula when both $M_1$ and $M_2$ are light mesons is
\begin{widetext}
\begin{eqnarray}
\langle M_1 M_2 | O_i | B \rangle & = & F_0^{B \to M_1} f_{M_2}
\int dx \, T_{M_2 ,i}^I \phi_{M_2} (x) \; + \; (1 \leftrightarrow
2) \nonumber \\* & + & f_B f_{M_1} f_{M_2} \int dz \, dy \, dx \,
T_i^{II}(x,y,z) \phi_B(z) \phi_{M_1}(y) \phi_{M_2}(x) \,  ,
\end{eqnarray}
\end{widetext}
where $\phi$ are the leading-twist light-cone distribution
amplitudes and the integration is over the momentum fractions
inside the mesons, $T_i^I$ includes tree diagrams plus corrections
(hard gluon exchanges and penguins) and $T_i^{II}$ expresses the
hard gluon exchange with the spectator. For $T_i^I =1$ and
$T_i^{II}=0$, we recover the naive factorization. The meson wave
functions will be an important source of SU(3) breaking. For the
light mesons, we have twist-2 and twist-3 distribution amplitudes
respectively defined in the following bilocal operator matrix
elements:
\begin{widetext}
\begin{eqnarray}
\langle M(p) | \bar{q}(z_2) \gamma_{\mu} \gamma_5 q(z_1)|0 \rangle
& = & - \, f_M p_{\mu} \int_0^1 dx \; e^{i(x p \cdot z_2 + \bar{x}
p \cdot z_1)} \phi(x) \nonumber \\* \langle M(p) | \bar{q}(z_2) i
\gamma_{5} q(z_1)|0 \rangle & = & f_M \mu_M \int_0^1 dx \; e^{i(x
p \cdot z_2 + \bar{x} p \cdot z_1)} \phi_p(x) \, ,
\end{eqnarray}
\end{widetext}
where $\mu_M$ is expressed in terms of the quark masses as $\mu_M
=m_M^2/(m_1+m_2)$ and $\bar{x} =1-x$. In the momentum space, the
light-cone projector operator of a light pseudoscalar meson
described by both the twist-2 and twist-3 amplitudes is:
\begin{equation}
\Phi (M) = \frac{i f_M}{4 N_c} \left \lbrace \hat{p} \gamma_5
\phi(x) - \mu_M \, \gamma_5 \phi_p(x) \right \rbrace ,
\end{equation}
where $\hat{p} = p \cdot \gamma$.

We notice that in $a_1, \, a_2 , \, a_3 , \, a_4 , \, a_5, \, a_7
, \, a_9 , \, a_{10}$, where we have $(V-A)(V \pm A)$, only the
twist-2 amplitude is taken, while in $a_6 , \, a_8$ (the terms
proportional with $r = 2 \mu/m_b$) only the twist-3 amplitude must
be considered. The operators $a_6$ and $a_8$ are important for
penguin-dominant $B$ decays.

The twist-2 distribution amplitude, $\phi(x)$, has the following
expansion in Gegenbauer polynomials [2]
\begin{eqnarray}
\phi (x) & = & 6x(1-x)[ 1+ \alpha_1 C^{(3/2)}_1(2x-1) \nonumber
\\* & + & \alpha_2C^{3/2}_2(2x-1) + ...] ,
\end{eqnarray}
with $C^{3/2}_1(u) = 3 u$ and $C^{3/2}_2(u) = (3/2)(5u^2-1)$, and
is different for $\pi$ and $K$. For $\pi$, the distribution in $x$
must be even because the $u$ and $d$ quarks have negligible masses
and their distributions inside the pion are symmetric. This
dictates $\alpha_1^\pi = 0$. The coefficient $\alpha^\pi_2$ is
estimated to be $0.25 \pm 0.15$. For K, the $u$ (or $d$) and $s$
quarks inside the kaon are different, leading to an asymmetry in
the $x$ distribution. So a non-zero value for $\alpha^K_1$ is
needed and it is estimated to be $0.05$, while $\alpha^K_2 =
\alpha^\pi_2$. The leading twist distribution amplitudes, valid
for $\mu \to \infty$, are $\phi(x) =6x(1-x)$ and $\phi_p(x) =1$.

The $B$ meson distribution between the heavy quark and light
antiquark is described by
\begin{equation}
\phi_B = {\cal N} \frac{z^2 (1-z)^2}{(a^2 (1-z) + z^2)^2} \, ,
\end{equation}
where the parameter $a$ is related to the position of the maximum
of the amplitude and is very small $a \in [0.05 \div 0.1]$.
However, since the momentum is almost carried by the heavy quark,
one may work with a strongly peaked function around $z_0 =
\lambda_B/m_B \approx 0.066 \pm 0.029$, for $\lambda_B =0.35 \pm
0.15$ GeV.

In the this approach, [3], the $a_i$ coefficients,
\begin{widetext}
\begin{eqnarray}
a_1(M_1 M_2) & = & C_1 + {C_2 \over N_c} \left[ 1 + {C_F \alpha_s
\over 4 \pi} (V_{M_2} +  H) \right],
\nonumber\\
a_2(M_1 M_2) & = & C_2 + {C_1 \over N_c} \left[ 1 + {C_F \alpha_s
\over 4 \pi} (V_{M_2} +  H) \right],
\nonumber\\
a_3(M_1 M_2) & = & C_3 + {C_4 \over N_c} \left[ 1 + {C_F \alpha_s
\over 4 \pi} (V_{M_2} +  H) \right],
\nonumber\\
a^p_4(M_1 M_2) & = & C_4 + {C_3\over N_c} \left[ 1 + {C_F \alpha_s
\over 4 \pi} (V_{M_2} + H) \right] + {C_F \alpha_s \over 4\pi N_c}
P^p_{M_2,2},
\nonumber\\
a_5(M_1 M_2) & = & C_5 + {C_6 \over N_c} \left[ 1 + {C_F \alpha_s
\over 4 \pi} (-12-V_{M_2} - H) \right],
\nonumber\\
a^p_6(M_1 M_2) & = & C_6 + {C_5\over N_c} \left( 1 - 6 {C_F
\alpha_s \over 4 \pi} \right) + {C_F\alpha_s \over 4\pi N_c}
P^{p}_{M_2, 3},
\nonumber\\
a_7(M_1 M_2) & = & C_7 + {C_8 \over N_c} \left[ 1 + {C_F \alpha_s
\over 4 \pi} (-12-V_{M_2} - H) \right],
\nonumber\\
a^p_{8}(M_1 M_2) & = & C_8 + {C_7 \over N_c} \left( 1 - 6{C_F
\alpha_s\over 4 \pi} \right) + {\alpha\over 9\pi N_c}
P^{p,EW}_{M_2, 3},
\nonumber\\
a_9(M_1 M_2) & = & C_9 + {C_{10} \over N_c} \left[ 1 + {C_F
\alpha_s \over 4 \pi} (V_{M_2} +  H) \right],
\nonumber\\
a^p_{10}(M_1 M_2) & = & C_{10} + {C_9\over N_c} \left[ 1 + {C_F
\alpha_s \over 4 \pi} (V_{M_2} + H) \right] + {\alpha \over 9 \pi
N_c} P^{p,EW}_{M_2, 2},
\end{eqnarray}
\end{widetext}
where $C_F = (N^2_c-1)/2N_c$, include the vertex, the hard gluon
exchange with the spectator and the penguin contributions, at $\mu
=m_b$, defined as
\begin{widetext}
\begin{eqnarray}
V_M & = & - \, 18 + \int^1_0 dx g(x) \phi_M(x), \nonumber\\*
P^p_{M,2} & = & C_1 \left[ \frac{2}{3} + G_M(s_p) \right] + C_3
\left[ \frac{4}{3} + G_M(0) + G_M(3) \right] \nonumber\\* & + &
(C_4+C_6) \left[ (n_f-2) G_M(0) + G_M (s_c) + G_M(1) \right] - \,
2 C_{8g}^{eff} \int^1_0 {dx\over \bar{x}} \phi_M(x), \nonumber \\*
P^{p,EW}_{M,2} & = & (C_1 + N_c C_2) \left[ \frac{2}{3} + G_M(s_p)
\right] - 3 C_{7\gamma}^{eff} \int^1_0 {dx\over \bar{x} }
\phi_M(x), \nonumber \\* P^p_{M,3} & = & C_1 \left[ \frac{2}{3} +
\hat{G}_M(s_p) \right] + C_3 \left[ \frac{4}{3} + \hat{G}_M(0) +
\hat{G}_M(1) \right] \nonumber\\* & + & (C_4+C_6) \left[ (n_f-2)
\hat{G}_M(0) + \hat{G}_M(s_c) + \hat{G}_M(1) \right] - 2
C_{8g}^{eff} \, , \nonumber\\
P^{p,EW}_{M,3} & = & (C_1 + N_c C_2) \left[ \frac{2}{3} +
\hat{G}_M(s_p) \right] - 3 C_{7\gamma}^{eff} \, ,
\nonumber\\
H & = & \frac{4 \pi^2}{N_c} \frac{f_B f_{M_1}}{m_B \lambda_B
F^{B\to M_1}_0(0)} \, \int^1_0 {dx\over \bar{x}} \, \phi_{M_2}(x)
\int^1_0 {dy \over \bar{y}} \left[ \phi_{M_1}(y) + \frac{2
\mu_{M_1}}{m_b} \frac{\bar{x}}{x} \phi^p_{M_1}(y) \right] ,
\end{eqnarray}
\end{widetext}
where the parameter $2\mu_M/m_b$ coincides with $r$ and $s_i  =
m_i^2/m_b^2$ are the mass ratios for the quarks involved in the
penguin diagrams, namely $s_u = s_d = s_s =0$ and $s_c =
(1.3/4.2)^2$.

Putting everything together, the amplitudes (8) get the explicit
form
\begin{widetext}
\begin{eqnarray}
A(\bar B^0 \to \pi^+ \pi^-) &=& - \, i {G_F \over \sqrt{2}} (m^2_B
- m^2_\pi) F^{B\to \pi}_0(m^2_\pi) f_\pi [V_{ub}V_{ud}^*
a_1(\pi\pi) \nonumber \\* &+& V_{pb}V_{pd}^*(a^p_4(\pi\pi)+
a^p_{10}(\pi\pi) + r_{\pi} (a_6^p(\pi\pi) + a_8^p(\pi\pi)))] \,  ,
\\
A(\bar B^0 \to \pi^+ K^-) &=& - \, i {G_F\over \sqrt{2}} (m^2_B -
m^2_\pi) F^{B\to \pi}_0(m^2_K) f_K [V_{ub}V_{us}^*a_1(\pi K )
\nonumber\\* &+& V_{pb}V_{ps}^*(a^p_4(\pi K)+ a^p_{10}(\pi K) +
r_K (a_6^p(\pi K) + a_8^p(\pi K)))] , \nonumber \\*
\end{eqnarray}
where $p$ is summed over $u$ and $c$, and consequently, the
relation (14) turns into, [6],
\begin{equation}
{\Delta^{\bar B^0}_{\pi^+\pi^-}\over \Delta^{\bar B^0}_{\pi^+
K^-}} \approx -{f^2_\pi\over f^2_K} \left[ \frac{1- 0.748
\alpha^\pi_1 - 0.109 \alpha^\pi_2 - 0.0013 H_{\pi \pi} }{ 1- 0.748
\alpha^K_1  - 0.109 \alpha^K_2 - 0.0013 H_{\pi K}} \right],
\end{equation}
\end{widetext}
pointing out the following SU(3) breaking effects: the difference
in the decay constants and form factors, the difference in the
$\alpha_1$ and $\alpha_2$ coefficients that appear in the twist-2
distribution amplitudes (19) and the $H_{\pi\pi}$ and $H_{\pi K}$
contributions (defined in (22)).

The decay amplitudes for $\bar B_s^0 \to K^+ \pi^-$ and $\bar
B_s^0 \to K^+ K^-$ can be obtained by using the appropriate
transition form factor $F^{B_s\to K}_0$ and by changing
$1/m^2_B\lambda_B$ to $1/m^2_{B_s} \lambda_{B_s}$ in $H_{M_1M_2}$.
One gets the same expression, (25), and thus we have come to the
following relation:
\begin{eqnarray}
{\Delta^{\bar B^0}_{\pi^+\pi^-}\over \Delta^{\bar B^0}_{\pi^+
K^-}} \, \approx \, {\Delta^{\bar B^0_s}_{K^+ \pi^-}\over
\Delta^{\bar B^0_s}_{K^+ K^-}} \, .
\end{eqnarray}
These example shows that important SU(3) breaking effects arise
from the light-cone distributions of mesons in addition to those
already present in the decay constants. These effects can only be
estimated with large uncertainty because the parameters
$\alpha^P_{1,2}$ are not well determined at present. Using the
currently allowed ranges we find,
\begin{equation}
A_{CP}(\pi^+ \pi^- ) \approx - \left( 3.1^{+1.9}_{-0.9} \right)
A_{CP}(\pi^+ K^-)\; ,
\end{equation}
which can also be used to test the SM and the IFM to some extent.

However, relations which are independent of $\alpha_{1,2}^i$
parameters and decay constants, such as (26), are more reliable
since they do not receive the main SU(3) breaking corrections that
we have investigated.

\subsection{$B \to PV$ Decays}

When the vector meson is factored out, as in
\[
\bar{B}^0 \to \pi^+ \rho^- \, , \; \bar{B}^0_s \to K^+ K^{*-} \, ,
\]
\[
\bar{B}^0_s \to K^+ \rho^- \, , \; \bar{B}^0 \to \pi^+ K^{*-}
\; ,
\]
the decay amplitudes can be obtained by replacing the $r_K$ factor
with $r_K^* = \frac{2 m_{K^*}}{m_b} \,
\frac{f^{\perp}_{K^*}}{f_{K^*}} \approx 0.3$ (and similarly for
$r_{\rho}$), and by removing the penguin terms $P^{p,EW}_{M_2,3}$
in the expressions for $a_6$ and $a_8$ (the vector meson is
described only by a twist-2 distribution amplitude). With all
these taken into account, we get, [6],
\begin{eqnarray}
& & \frac{\Delta^{\bar{B}^0}_{\pi^+ \rho^-}}{\Delta^{
\bar{B}^0_s}_{K^+ K^{*-}}} \approx - \, {m_{B}\over
m_{B_s}}\frac{f_{\rho}^2}{f_{K^*}^2} \nonumber \\* & & \times
\left( \frac{F_1^{B \to \pi}}{F_1^{B_s \to K}} \right)^2 \frac{1 -
1.25 \alpha_1^{\rho} - 0.18 \alpha_2^{\rho}}{1 - 1.25
\alpha_1^{K^*} -0.18 \alpha_2^{K^*}} \; .
\end{eqnarray}
Using the central values of the ranges
\[
\alpha_1^{\rho} =0 \, , \; \alpha_2^{\rho}=0.15 \, , \]
\[
\alpha_1^{K^*}=0.04 \, , \; \alpha_2^{K^*}=0.10
\]
and taking $f_{\rho} \approx 0.96 f_{K^*}$, we find,
\begin{eqnarray}
\frac{\Delta^{\bar{B}^0}_{\pi^+ \rho^-}}{
\Delta^{\bar{B}^0_s}_{K^+ K^{*-}}} & \approx & - \, 0.95 \left(
\frac{F_1^{B \to \pi}}{F_1^{B_s \to K}} \right)^2 \; , \nonumber
\\* \frac{\Delta^{\bar{B}^0_s}_{K^+ \rho^-}}{
\Delta^{\bar{B}^0}_{\pi^+ K^{*-}}} & \approx & - \, 0.95 \left(
\frac{F_1^{B_s \to K}}{F_1^{B \to \pi}} \right)^2.
\end{eqnarray}

When the meson that picks up the spectator is a vector, as for
example in
\[
\bar B^0 \to \rho^+ \pi^- \, , \; \bar B^0_s \to K^{*+} K^- \, ,
\]
\[
\bar B^0 \to \rho^+ K^- \, , \; \bar B_s^0 \to K^{*+} \pi^- \;
,
\]
the corresponding decay amplitudes can be obtained by replacing
the form factor $F^{B\to P}_0$ with $A^{B\to V}_0$ and $r$ with
$-r$. In this case, the SU(3) breaking is large and estimates are
unreliable since the analogue of (29), [6],
\begin{eqnarray}
& & \frac{\Delta^{\bar B^0}_{\rho^+ \pi^-}}{ \Delta^{\bar
B^0_s}_{K^{*+} K^-}} \, \approx \, - \, {m_{B}\over m_{B_s}}
\frac{f_{\pi}^2}{f_K^2} \nonumber \\* & & \times \left(
\frac{A_0^{B \to \rho}}{A_0^{B_s \to K^*}} \right)^2 \frac{1 + 110
\alpha_1^{\pi} + 15.5 \alpha_2^{\pi}}{1+ 110 \alpha_1^K + 15.5
\alpha_2^K}\;,
\end{eqnarray}
contains large coefficient of $\alpha_1$ in both the numerator and
denominatorm making a prediction for this asymmetry impossible
within this framework. On the other hand, this provides an
opportunity to constrain (or even to determine) $\alpha_1^K$ when
the ratio in (30) is measured.

\subsection{$B^- \to K^- \eta^{\prime}$}

Let us turn now to the $B^- \to \eta^{\prime} K^-$ decay which has
become of a real interest after CLEO announced its large numerical
value $BR(B^- \to \eta^{\prime}K^- ) = ( 6.5^{+1.5}_{-1.4} \pm 0.9
) \times 10^{-5}$, which could not be explained by the existent
theoretical models. As improved measurements followed, providing
even larger values, $(80^{+10}_{-9} \pm 7) \times 10^{-6}$ (CLEO)
and $(76.9 \pm 3.5 \pm 4.4) \times 10^{-6}$ (BaBar), inclusion of
new contributions for accommodating these data has quickly become
a real theoretical challenge.

The relevant decay amplitude for $ B^- \to \eta^{\prime} K^-$ is,
[4],
\begin{widetext}
\begin{eqnarray}
A(B^- \to \eta^{\prime} K^-) &=& - \, i {G_F \over \sqrt{2}}
(m^2_B - m^2_{\eta^{\prime}}) F^{B\to \eta^{\prime}}_0(m^2_K) f_K
\left[ V_{ub} V_{us}^* a_1(X) \right. \nonumber \\* & & \left. +
\, V_{pb}V_{ps}^* \left( a^p_4(X)+ a^p_{10}(X) + r_K (a_6^p(X) +
a_8^p(X)) \right) \right] \nonumber \\* &-& i {G_F \over \sqrt{2}}
(m^2_B - m^2_K) F^{B\to K}_0(m^2_{\eta^{\prime}})
f^u_{\eta^{\prime}} \left[ V_{ub}V_{us}^* a_2(Y) + V_{pb}V_{ps}^*
\left[ \left( a_3(Y)-a_5(Y) \right) ( 2 + \sigma ) \right. \right.
\nonumber \\* & & \left. \left. + \left[ a_4^p (Y) - \frac{1}{2}
a_{10}^p (Y) + r^{\prime} \left( a_6^p(Y) - \frac{1}{2} a_8^p (Y)
\right) \right] \sigma  + \frac{1}{2} \left( a_9(Y) - a_7(Y)
\right) (1 - \sigma ) \right] \right] ,
\end{eqnarray}
\end{widetext}
where $X = \eta^{\prime} K$ and $Y = K \eta^{\prime}$, $r^{\prime}
= 2m^2_{\eta^{\prime}}/(m_b-m_s)(2m_s)$ and $\sigma =
f^s_{\eta^{\prime}}/f^u_{\eta^{\prime}}$. As it can be noticed,
the coefficients $a_i$ are different for the $X$ and $Y$ final
states, since they depend on the twist-2 and twist-3 wave
functions of the $M_2$ meson, except for the hard contribution
where the wave functions for both $M_1$ and $M_2$ are involved.

The twist-2 distribution amplitude $\phi_K(x)$ has the usual
expansion in Gegenbauer polynomials, while the corresponding
twist-3 amplitude, $\phi^p_K$, is 1. In what it concerns the
physical states $\eta$ and $\eta^{\prime}$, these are mixtures of
SU(3)-singlet and octet components $\eta_0$ and $\eta_8$ and the
corresponding decay constants, in the two-angle mixing formalism,
are given by, [1], $f^u_{\eta^{\prime}} = 63.5$ MeV,
$f^s_{\eta^{\prime}} = 141$ MeV, while the relevant form factor
for the $B \to \eta^{\prime}$ transition is $ F_0^{B \to
\eta^{\prime}} =  0.137$. Even the $\eta^{\prime}$ flavor singlet
meson has a gluonic content, [10], which could bring a
contribution to the wave function, this is supposed to be small
and therefore we employ, in the calculation of
$V_{\eta^{\prime}}$, $P^p_{\eta^{\prime},2}$ and
$P^{p,EW}_{\eta^{\prime},2}$ in $a_i(Y)$, only the leading twist-2
distribution amplitude $\phi_{\eta^{\prime}} = 6 x \bar{x}$. Also,
since the twist-3 quark-antiquark distribution amplitude do not
contribute, due to the chirality conservation, the penguin parts
in $a_6^p(Y)$ and $a_8^p(Y)$ are missing.

In IFM, we get for the $B^- \to K^- \eta^{\prime}$ decay the
numerical value $Br(B \to K \eta^{\prime}) = 3.65 \cdot 10^{-5}$
which is comparable to other theoretical estimations, [1, 4, 10],
but is only half of the averaged experimental data, suggesting
that one has to incorporate take into account new contributions in
order to increase the $Br(B \to K \eta^{\prime})$ numerical
values.

In this respect, we have employed ing the Minimal Supersymmetric
Standard Model (MSSM), by adding to the effective SM Hamiltonian
the following SUSY contribution
\begin{widetext}
\begin{equation}
H^{SUSY}_{7-8} \, = \,  - \, i \, {G_F \over \sqrt{2}} (V_{ub}
V_{us}^* +V_{cb} V_{cs}^*) \left( c_{8g}^{SUSY} O_{8g} + c_{7
\gamma}^{SUSY} O_{7 \gamma} \right) ,
\end{equation}
expressed in terms of the gluon and photon operators:
\begin{equation}
O_{8g} \, = \, \frac{g_s}{8 \pi^2} \, m_b \bar{s} \sigma_{\mu \nu}
(1+\gamma_5 ) G^{\mu \nu} b \; , \; \; O_{7 \gamma} \, = \,
\frac{e}{8 \pi^2} \, m_b \bar{s} \sigma_{\mu \nu} (1+\gamma_5 )
F^{\mu \nu} b \, .
\end{equation}
The Wilson coefficients are given by, [5],
\begin{eqnarray}
c_{8g}^{SUSY} (M_{SUSY}) & = & - \, \frac{\sqrt{2} \pi
\alpha_s}{G_F (V_{ub} V_{us}^* + V_{cb} V_{cs}^*) m_{\tilde{g}}^2}
\, \delta^{bs}_{LR} \, \frac{m_{\tilde{g}}}{m_b} \, G_0(x) \, ,
\nonumber \\
c_{7 \gamma}^{SUSY} (M_{SUSY})  & = & - \, \frac{\sqrt{2} \pi
\alpha_s}{G_F (V_{ub} V_{us}^* + V_{cb} V_{cs}^*) m_{\tilde{g}}^2}
\, \delta^{bs}_{LR} \, \frac{m_{\tilde{g}}}{m_b} \, F_0(x) \, ,
\end{eqnarray}
where
\begin{eqnarray}
G_0 (x) & = & \frac{x}{3(1-x)^4} \, \left[ 22-20x-2x^2+16 x \ln(x)
-x^2 \ln (x) + 9 \ln (x) \right] ,
\nonumber \\
F_0 (x) & = & - \; \frac{4x}{9(1-x)^4} \, \left[ 1+4x-5x^2+4 x
\ln(x) + 2 x^2 \ln (x) \right] .
\end{eqnarray}
\end{widetext}
In the above relations, $x=m_{\tilde{g}}^2 / m_{\tilde{q}}^2$,
with $m_{\tilde{g}}$ being the gluino mass and $m_{\tilde{q}}$ an
average squark mass, while the factor $\delta^{bs} = \Delta^{bs}/
m_{\tilde{q}}^2$, where $\Delta^{bs}$ are the off-diagonal terms
in the sfermion mass matrices, comes from the expansion of the
squark propagator in terms of $\delta$, for $\Delta \ll
m_{\tilde{q}}^2$. In principle, the dimensionless quantities
$\delta^{bs}$, measuring the size of flavor changing interaction
for the $\tilde{s} \tilde{b}$ mixing, are present in all the SUSY
corrections to the Wilson coefficients and they are of four types,
depending on the $L$ or $R$ helicity of the fermionic partners.
However, one finds that $\lbrace c_i^{SUSY}
\rbrace_{i=\overline{3,6}}$, for $M_{SUSY} = m_{\tilde{q}}= 500$
GeV and $x \approx 1$, do not bring any significant contribution
to the branching ratio. The situation looks different in what it
concerns the SUSY Wilson coefficients (34) that, when
$m_{\tilde{g}}$ is of order of few hundred GeV, will dominate the
SM ones.

Thus, we replace the Wilson coefficients $c_{8g}^{eff}$ and $c_{7
\gamma}^{eff}$, by the total quantities
\begin{eqnarray}
c_{8g}^{total} [x, \delta] & = & c_{8g}^{eff} + c_{8g}^{SUSY}
(m_b) \; , \nonumber \\* c_{7 \gamma}^{total} [x, \delta] & = &
c_{7 \gamma}^{eff} + c_{7 \gamma}^{SUSY} (m_b) \; ,
\end{eqnarray}
where $c^{SUSY} (m_b)$ have been evolved from $M_{SUSY} =
m_{\tilde{g}}$ down to the $\mu =m_b$ scale. For $m_{\tilde{q}} =
500$ GeV, $m_{\tilde{g}} = m_{\tilde{q}}$ and $\delta^{bs}_{LR}
\equiv \rho e^{i \varphi}$, the total branching ratio can be
expressed in terms of the parameters $\rho$ and $\varphi$ as
\begin{widetext}
\begin{equation}
BR^{total} \, = \, 10^{-5} \left( 3.65 + 447 \rho \cos \varphi +
13670 \rho^2 + 13.78  \rho \sin \varphi \right) ,
\end{equation}
\end{widetext}
pointing out, besides the (IFM)-value $3.65 \times 10^{-5}$, the
SUSY contribution depending on $\rho$ and $\varphi$.

A detailed analysis of this formula, suggests that one should take
$\rho \in [0.005 , 0.01 ]$ and $\varphi \in [-3 \pi/4 , 3 \pi /4
]$, for accommodating the range within the two extreme
experimental data, $BR_{exp} (BaBar) = 7 \times 10^{-5}$ and
$BR_{exp} (CLEO) = 8 \times 10^{-5}$. For $\rho $ close to the
lowest limit of its interval, the predicted $BR^{total}$-values
lie below the experimental data, while for $\rho$ moving to the
central value and $- \pi/4 \leq \varphi \leq \pi /4$, one gets
$BR_{total} \in [ 7 \times 10^{-5} , 8 \times 10^{-5} ]$.

\begin{acknowledgments}
The authors gratefully acknowledge the kind hospitality and
fertile environment of the University of Oregon where this work
has been carried out. Special thanks go to Damir Becirevic and
Vladimir Braun for useful discussions and fruitful suggestions.
This work is supported by the CNCSIS Type A Grant, Code 1433/2007.
\end{acknowledgments}

\end{document}